\documentclass{PoS}

\def\tr{\,\hbox{tr}\,}
\title{Dynamical quark mass generation}

\ShortTitle{Dynamical quark mass}

\author{\speaker{R.\ Frezzotti}
\\
        {Dipartimento di Fisica, Universit\`a di Roma {\it Tor Vergata}\\ \small and INFN, Sezione di Roma Tor Vergata \\ \small Via della Ricerca Scientifica, 00133 Roma, Italy}\\
        E-mail: \email{frezzotti@roma2.infn.it}}

\author{G.C.\ Rossi\\
       {Dipartimento di Fisica, Universit\`a di Roma {\it Tor Vergata}\\ \small and INFN, Sezione di Roma Tor Vergata \\ \small Via della Ricerca Scientifica, 00133 Roma, Italy}\\
      E-mail: \email{rossig@roma2.infn.it}}

\abstract{Taking inspiration from lattice QCD results, we argue that a non-perturbative mass term for fermions can be generated as a consequence of the dynamical phenomenon of spontaneous chiral symmetry breaking, in turn triggered by the explicitly breaking of chiral symmetry induced by the critical Wilson term in the action. In a pure lattice QCD-like theory this mass term cannot be separated from the unavoidably associated linearly divergent contribution. However, if QCD with a Wilson term is enlarged to a theory where also a scalar field is present, coupled to a doublet of SU(2) fermions via a Yukawa interaction, then in the phase where the scalar field takes a non-vanishing (large) expectation value, a dynamically generated and ``naturally'' light fermion mass (numerically unrelated to the expectation value of the scalar field) is seen to emerge, at a critical value of the Yukawa coupling where the symmetry of the model is maximally enhanced.
}

\FullConference{31st International Symposium on Lattice Field Theory - LATTICE 2013\\
		July 29 - August 3, 2013\\
		Mainz, Germany}

\begin{document}

\section{Introduction and outlook}
\label{sec:INTRO}

In this contribution we argue that in non-Abelian 
gauge theories with chiral symmetries broken at the UV cutoff by Wilson-like terms the dynamics of spontaneous chiral symmetry breaking (S$\chi$SB) - triggered in the critical limit by the residual explicit chiral breaking - generates a dynamical mass for fermions.
If one can solve, as we are going to show in a simple model including QCD, the ``naturalness'' problem~\cite{THOOFT} associated to the need of ``fine tuning'' the 
parameter controlling the recovery of chiral symmetry, this road may lead to a viable non-perturbative (NP) analog of the Higgs mechanism for mass generation~\cite{FRNEW}. In such a framework electroweak interactions can be naturally introduced. 
If a superstrong interaction at the TeV scale is also introduced, one can set up a model 
where mass hierarchy and the flavour properties of the Standard Model (recovered as the low energy theory) are understood and arise in a natural way.

\section{Inspiration and numerical evidence from lattice QCD}
\label{sec:LQCD}

As is well known, in lattice QCD (LQCD) with Wilson fermions~\cite{WIL} quark mass renormalization requires the subtraction of a linearly divergent counter-term, $m_{cr}\bar q q$ ($q $ being the $N_f$-flavour quark field), arising because the Wilson term in the lattice Lagrangian explicitly breaks chiral symmetry.
In general $m_{cr}$ will have a formal small-$a$ expansion of the kind 
\vspace{-0.1cm}
\begin{equation}
m_{cr} = \frac{c_0}{a}+c_1 \Lambda_{QCD}+c_2 a\Lambda_{QCD}^2+{\mbox{O}}(a^2)\, .
\label{CRM1}
\vspace{-0.1cm}
\end{equation}
Eq.~(\ref{CRM1}) suggests that, if we could set the mass parameter, $m_0$, in the lattice fermion action just equal to the linearly divergent term in~(\ref{CRM1}), then the $c_1 \Lambda_{QCD}$ contribution (if non-zero) would play the role of a quark mass in the renormalized chiral Ward--Takahashi Identities (WTIs). To make use of this remark for NP mass generation one has to answer positively the following questions. 

1) Are there numerical indications for the existence of a term like the second one in the r.h.s.\ of~(\ref{CRM1}) in actual LQCD simulation data? 2) Do we understand its dynamical origin? 
3) Is it possible to disentangle a (small) NP fermion mass from the much larger (perturbative) effect that comes along with it when chiral symmetry is broken at a high scale? 
\begin{figure}[htbp]
\centerline{\includegraphics[scale=0.30,angle=0]{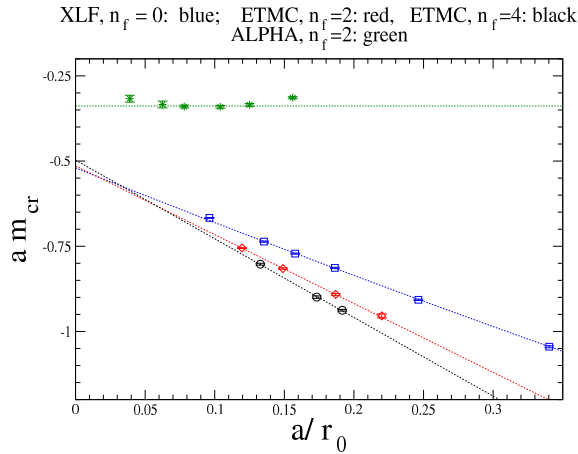}}
\vspace*{-0.5cm}
\caption{\small{The critical value of $am_0$ in Wilson LQCD simulations as a function of $a/r_0$.}}
\label{fig:BAMBI}
\end{figure}

To answer question 1), in Fig.~\ref{fig:BAMBI} we present a compilation of LQCD data showing the behaviour of $am_{cr}$, which is the value of the $am_0$ Lagrangian parameter at which $am_{PCAC}$ vanishes,
as a function of $a/r_0$ (here $r_0$ denotes the Sommer scale).
We show four sets of data taken from refs.~\cite{Jansen:2005kk}, \cite{Baron:2009wt,Blossier:2010cr}, \cite{Baron:2010bv} and~\cite{DellaMorte:2004bc}. The three lower sets of points correspond to measurements of $am_{cr}$ carried out at maximal twist 
using the Wilson twisted mass regularization of LQCD 
in the quenched ($N_f=0$) approximation (blue squares~\cite{Jansen:2005kk}), with $N_f=2$ dynamical flavours (red diamonds~\cite{Baron:2009wt,Blossier:2010cr}) and with $N_f=4$ dynamical flavours (black circles~\cite{Baron:2010bv}). The green stars correspond to Wilson clover-improved~\cite{Sheikholeslami:1985ij} data with $N_f=2$ dynamical flavours~\cite{DellaMorte:2004bc} obtained
with Schr\"odinger functional boundary conditions. For the sake of Fig.~\ref{fig:BAMBI} we have taken $r_0=0.45$~fm~\cite{Baron:2009wt,Blossier:2010cr}.

In the present notations the intercept of the fitted line through the data is the $c_0$ coefficient of~(\ref{CRM1}), while its slope, $c_1 \Lambda_{QCD}\times r_0$, is the quantity of interest. Indeed, the points of refs.~\cite{Jansen:2005kk}, \cite{Baron:2009wt,Blossier:2010cr}, \cite{Baron:2010bv} all exhibit a nice linear behaviour (with a mild $N_f$ dependence) in a wide $a/r_0$ window, 
which allows identifying a non-vanishing $c_1 \Lambda_{QCD}$ slope taking values~\footnote{A word of caution is in order here. The quoted values of $d(am_{cr})/ d (ar_0^{-1})$ are only indicative, as strictly speaking there isn't a mathematically rigorous way to identify an $a/r_0$ range where one can consider numerically negligible both the logarithmic $a$-behaviour of the gauge coupling upon renormalization (determining the behaviour of $am_{cr}$ as $a\to 0$) and the higher order lattice artifacts that become dominant at large enough values of $a$.}
in the range 700 to 1000~MeV.

The Schr\"odinger functional data of ref.~\cite{DellaMorte:2004bc} are, instead, pretty flat implying that the $c_1$ coefficient is very small. As we shall argue in the next section, 
a non-zero slope is related to the O($a$) Wilson-like term 
in the Symanzik low energy effective Lagrangian (SLEL)~\cite{SYM}. The presence of the non-perturbatively tuned clover-term~\cite{Sheikholeslami:1985ij} in the lattice Lagrangian employed in ref.~\cite{DellaMorte:2004bc} 
effectively kills (the interesting NP effects originating from) the $d=5$ SLEL operator~\cite{Luscher:1996sc}.

\section{The dynamical origin of the $c_1 \Lambda_{QCD}$ term}
\label{sec:DYNOR}

\begin{figure}[htbp]
\centerline{\includegraphics[scale=0.40,angle=0]{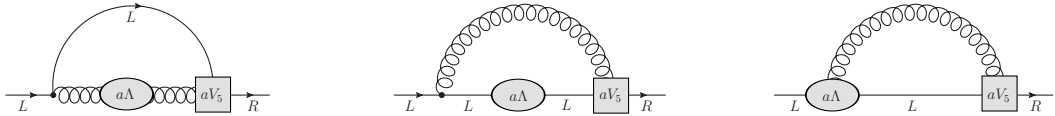}}
\caption{\small{Typical lowest order ``diagrams'' giving rise to dynamically 
generated quark mass 
terms (L and R are quark-helicity labels). 
The grey blob represents the non-perturbative 
$a\Lambda_{QCD}\alpha_s$ effect in eqs.~(3.2).}}
\label{fig:FIG6}
\end{figure}

The $c_1 \Lambda_{QCD}$ term in~(\ref{CRM1}) has its origin in a delicate 
interplay 
between O($a$) corrections to 
quark and gluon propagators and vertices ensuing from the spontaneous breaking of chiral symmetry, and the quadratic divergence of the loop integration in diagrams where one Wilson term vertex is inserted. 
Typical (lowest order in $g_s^2$) correlators where this occurs are 
depicted in Fig.~\ref{fig:FIG6}, where the grey blob is a NP O($a$) correction to
the gluon or (the helicity-preserving components of) the quark propagator and the gluon-quark-quark vertex, and $aV_5$ is the derivative vertex from the Wilson term.
Such peculiar O($a$) corrections arise from NP contributions 
in the SLEL expansion of the relevant (gauge fixed) lattice 
correlators for $m_0 \to m_{cr}$, e.g.\
\begin{eqnarray}
\hspace*{-.6cm} &&\langle O(x,x',...) \rangle\Big{|}^L = 
\langle O(x,x',...)  \rangle\Big{|}^C - 
a  \langle O(x,x',...) \int d^4 z L_5(z) \rangle\Big{|}^C + 
{\rm O}(a^2) \, ,\label{ALC1SYM}\\
\hspace*{-.6cm} && O(x,x',...) \; \Leftrightarrow  \;
A_\mu^b(x) A_\nu^c(x') \, , \;\;  q_{L/R}(x) \bar q_{L/R}(x') \, ,
\;\;  q_{L/R}(x) \bar q_{L/R}(x') A_\mu^b(y) \, , \nonumber 
\end{eqnarray}
where $L_5$ is the $d=5$ SLEL operator, which breaks chirality.
The label $|^C$ is to remind that the r.h.s.\  correlators are taken in continuum (renormalized) QCD.
The key remark about these expansions is that the O($a$) correlators in~(\ref{ALC1SYM}) can be non-zero only due to the phenomenon of S$\chi$SB. 
From these (amputated) correlators by using symmetry and dymensional arguments one reads off the NP contributions to quark and gluon propagators and vertices, 
namely (in continuum-like notations)
\begin{eqnarray}
\hspace{-1.cm}&&\Delta G_{\mu\nu}^{bc}(k) \Big{|}^L = - a\Lambda_{QCD} \,
\alpha_s(\Lambda_{QCD}) \, \delta^{bc}\,\frac{ \delta_{\mu\nu}-k_\mu k_\nu/k^2}{k^2}\, 
 \,f_{AA}\Big{(}\frac{\Lambda_{QCD}^2}{k^2}\Big{)}  \, ,\nonumber
\\
\hspace{-1.cm}&&\Delta S_{LL/RR}(k) \Big{|}^L = - a\Lambda_{QCD} \,\alpha_s(\Lambda_{QCD}) 
\,  \frac{i k_\mu  (\gamma_\mu)_{LL/RR} }{k^2} \,f_{q \bar q}\Big{(}\frac{\Lambda_{QCD}^2}{k^2}\Big{)} \, ,\label{DELTAGL}\\
\hspace{-1.cm}&&\Delta \Gamma^{b,\mu}_{Aq\bar q}(k,\ell) \Big{|}^L 
= a\Lambda_{QCD} \,\alpha_s(\Lambda_{QCD}) 
\, ig_s \lambda^b \gamma_\mu \,f_{Aq \bar q}\Big{(}\frac{\Lambda_{QCD}^2}{k^2},\frac{\Lambda_{QCD}^2}{\ell^2},\frac{\Lambda_{QCD}^2}{(k+\ell)^2}\Big{)}\, , \nonumber
\end{eqnarray}
where the factor $\alpha_s(\Lambda_{QCD})$ comes from the fact that the gluon emitted from the $L_5$ vertex has to be absorbed somewhere in 
the diagram~\footnote{Since a soft quark line (where S$\chi$SB occurs) exits the $L_5$ vertex, $\Lambda_{QCD}$ is chosen as the scale of $\alpha_s$. The scale at which the gauge coupling is evaluated will be a key feature to understand the fermion mass hierarchy problem~\cite{FRNEW}.}. 
The scalar form factors $f_{AA}$, $f_{q\bar q}$ and $f_{Aq \bar q}$ are dimensionless functions 
depending on $\Lambda_{QCD}^2/({\mbox{momenta}})^2$ ratios. 
From Symanzik's analysis of lattice artifacts, $a$-expansions like those in eqs.~(\ref{ALC1SYM}) are expected to be valid for 
small 
values of  squared momenta  
compared to $a^{-2}$. Here we assume that the NP effects encoded in eqs.~(\ref{ALC1SYM})--(\ref{DELTAGL}) persist up to large (i.e.\ comparable to $a^{-1}$) momenta, and {\it conjecture} the asymptotic behaviour
\begin{eqnarray}
f_{AA}\Big{(}\frac{\Lambda_{QCD}^2}{k^2}\Big{)} \stackrel{k^2\to\infty}\longrightarrow  {h_{AA}}\, , \,\,\,\,\,
f_{q \bar q}\Big{(}\frac{\Lambda_{QCD}^2}{k^2}\Big{)} \stackrel{k^2\to\infty} \longrightarrow  h_{q\bar q}\, , \,\,\,\,\,
f_{Aq \bar q}\Big{(}\frac{\Lambda_{QCD}^2}{k^2},\frac{\Lambda_{QCD}^2}{\ell^2},\frac{\Lambda_{QCD}^2}{(k+\ell)^2}\Big{)}\stackrel{k^2,\ell^2,(k+\ell)^2\to\infty} \longrightarrow h_{q\bar q} \, ,\nonumber 
\end{eqnarray}
where $h_{AA}$ and $h_{q \bar q}$ are O(1) constants and the last two limits are relaterd by gauge invariance.  

The above relative O($a\Lambda_{QCD} \alpha_s$) corrections to propagators and
vertices 
generate O($\Lambda_{QCD} \alpha_s g_s^2$) corrections to the quark
self-energy, hence NP mass terms. Actually there are more relevant NP corrections 
besides those in eqs.~(\ref{DELTAGL}) and fig.~\ref{fig:FIG6}, i.e.\ corrections 
to the Wilson-term induced vertices and helicity-flip quark propagator components.
Based on LQCD symmetries, to leading order in $g_s^2$ (and $a$) all 
the NP terms can be {\it effectively reproduced} in PT using 
{\it ad hoc modified Feynman rules},  
namely those obtained adding to the LQCD Lagrangian (in continuum-like notations) the terms 
\begin{equation}
\Delta L |_{ad\,hoc} = a\Lambda_{QCD} \alpha_s
\Big{\{} h_{AA} \frac{1}{2} {\rm tr}(F F) + h_{q \bar q} (\bar q \gamma_\mu D_\mu q) +
h_{Wil}(-\frac{a}{2}r) (\bar q  D^2 q) \Big{\}}  \, .
\label{DL-PT-LQCD} \end{equation}
To see how a dynamical mass gets generated, consider the loop momentum counting of, say, the ``diagram'' in the central panel of fig.~\ref{fig:FIG6} in the $a\to 0$ limit. One has factors $a\Lambda_{QCD}\alpha_s(\Lambda_{QCD})k_\mu/k^2$ and $1/k^2$ from the NP contribution to the quark propagator and the standard gluon propagator, respectively, and a factor $ak_\mu$ from the derivative coupling of the Wilson vertex. Including the extra $g_s^2$ power from the gluon loop, one gets schematically a 
fermion mass term of the order
\begin{equation}
a\Lambda_{QCD}g^2_s\alpha_s(\Lambda_{QCD}) \int^{1/a} d^4 k \frac{k_\mu}{k^2}\frac{1}{k^2} ak_\mu \sim g^2_s\alpha_s(\Lambda_{QCD})\Lambda_{QCD}\, .
\label{CONS}
\end{equation}
Other ``diagrams'' give similar NP mass contributions yielding in eq.~(\ref{CRM1}) 
$c_1 \sim {\rm O}(g^2_s\alpha_s(\Lambda_{QCD}))$ to leading order in $g_s^2$.

\section{Light mass fermions with natural fine tuning: a toy model}
\label{sec:FTP}

Separation of ``large'' (infinite in LQCD) from ``small'' (finite up to logs) contributions in formulae like~(\ref{CRM1}) is only possible on the basis of some symmetry. Though absent in LQCD, this long sought for symmetry can be seen to exist
in some enlarged theory where besides gauge interactions, an SU(2) fermion doublet is coupled to a scalar field, $\Phi$, via a Yukawa interaction and a Wilson-like term. 
To be concrete let us consider the renormalizable toy-model ($b^{-1}=$ UV cutoff)
\begin{eqnarray}
\hspace{-1.cm}&&{\cal L}_{\rm{toy}}(Q,G,\Phi) \; = \; {\cal L}_{kin}(Q,G,\Phi)+{\cal V}(\Phi) + {\cal L}_{Yuk}(Q,\Phi) + {\cal L}_{Wil}(Q,G,\Phi)  \, ,\label{SULL} \\
\hspace{-1.cm}&&{\cal L}_{kin}(Q,G,\Phi) \; = \;  \frac{1}{4}F^{a \; G}_{\mu\nu}F^{a \; G}_{\mu\nu}+\bar Q_L\gamma_\mu {\cal D}^{G}_\mu Q_L+\bar Q_R\gamma_\mu{\cal D}^{G}_\mu\,Q_R+\frac{1}{2}{\tr}\big{[}\partial_\mu\Phi^\dagger\partial_\mu\Phi\big{]} \, , \label{LKIN} 
\nonumber \\
\hspace{-1.cm}&&{\cal L}_{Yuk}(Q,\Phi) \; = \;  \eta\,\big{(} \bar Q_L\Phi Q_R+\bar Q_R \Phi^\dagger Q_L\big{)} \label{LYUK} \, , \qquad{\cal V}(\Phi) \; = \;  \frac{\mu_0^2}{2}{\tr}\big{[}\Phi^\dagger\Phi\big{]}+\frac{\gamma_0}{4}\big{(}{\tr}\big{[}\Phi^\dagger\Phi\big{]}\big{)}^2\label{LPHI} \, , 
\nonumber \\
\hspace{-1.cm}&&{\cal L}_{Wil}(Q,G,\Phi) \; = \;  \frac{b^2}{2}\,\big{(}\bar Q_L{\overleftarrow{\cal D}}\,^{G}_\mu\Phi {\cal D}^{G}_\mu Q_R+\bar Q_R \overleftarrow{\cal D}\,^{G}_\mu \Phi^\dagger {\cal D}^{G}_\mu Q_L\big{)}
\nonumber \, .
\end{eqnarray}
Besides obvious symmetries, ${\cal L}_{\rm toy}$ is invariant under the (global) $\chi_L\times\chi_R$ transformations 
\begin{eqnarray}
\hspace{-1.3cm}&&\bullet\,\chi_L:\quad \tilde\chi_L\otimes (\Phi\to\Omega_L\Phi) 
\, ,\quad 
{\mbox {with}} \qquad \tilde\chi_L: Q_{L}\rightarrow\Omega_{L} Q_{L} \, , 
\bar Q_{L}\rightarrow \bar Q_{L} \Omega_{L}^\dagger \, ,
\quad \Omega_L\in {\mbox{SU}}(2)_{L}\label{CHIL}\\
\hspace{-1.3cm}&&\bullet\,\chi_R:\quad \tilde\chi_R\otimes (\Phi\to\Phi\Omega_R^\dagger) 
\, ,\quad 
{\mbox {with}}\qquad \tilde\chi_R: Q_{R}\rightarrow\Omega_{R} Q_{R} \, ,
\bar Q_{R}\rightarrow \bar Q_{R}\Omega_{R}^\dagger\, ,  
\quad \Omega_R\in {\mbox{SU}}(2)_{R}
\end{eqnarray}
but not under the ``chiral'' transformations $\tilde \chi_L\times\tilde\chi_R$ acting only on fermions. However, much like it happens with the critical mass in LQCD when chiral 
symmetry is recovered~\cite{Bochicchio:1985xa}, 
a critical value of the Yukawa coupling, $\eta_{cr}$, exists where the transformations 
$\tilde \chi_L\times\tilde\chi_R$ become up to O($b^2$) a symmetry of ${\cal L}_{\rm{toy}}$.
This is so because, ignoring possible NP effects, at $\eta_{cr}$ the $\tilde \chi$-breaking terms ${\cal L}_{Wil}$ and ${\cal L}_{Yuk}$ compensate each other to O($b^0$), and $\Phi$ decouples. Hence among Green functions involving quarks and gluons the same relations are implied by $\tilde \chi_L\times\tilde\chi_R $ and $\chi_L\times\chi_R$ invariances.

Moving beyond PT, always at $\eta_{cr}$, 
one has to consider the possibility of
dynamical breaking of the (approximate) $\tilde \chi_L\times\tilde\chi_R$ symmetry
(S$\tilde\chi$SB).  We can have two very different scenarios. In both cases we assume 
that the scalar mass is much larger than $\Lambda_s$, the RGI scale of the theory.

If $\mu_0^2$ is such that ${\cal V}(\Phi)$ has a single minimum with $\langle \Phi \rangle = 0$,  
${\cal L}_{Wil}$ and ${\cal L}_{Yuk}$ (both linear in $\Phi$) are
expected to provide no seed for S$\tilde\chi$SB, so the $\tilde \chi_L\times\tilde\chi_R$ symmetry is thus realized \`a la Wigner~\cite{FRNEW}. 

Physics is drastically different if $\mu_0^2$ is such that a double well potential develops. 
In this case it is convenient to expand the scalar field around its vacuum expectation value (vev) writing 
\begin{equation}
\Phi(x) = ({\rm{v}}+ \sigma(x)) 1_{2\times 2} + i \vec{\pi}(x) \vec{\tau} \, , 
\label{PHI-AROUND-V}
\end{equation}
with $\vec{\pi}$ a triplet of massless pseudoscalar Goldstone bosons and $\sigma$ a scalar 
of mass $m_\sigma = {\rm O}(v)\gg \Lambda_s$. 
As (ignoring $\Phi$ fluctuations) the $d=6$ term ${\cal L}_{Wil}$ looks much like the $d=5$ Wilson term in LQCD, 
we expect the $\tilde \chi$-breaking terms in ${\cal L}_{Wil}$ (and ${\cal L}_{Yuk}$) 
to now trigger dynamical S$\tilde \chi$SB~\cite{FRNEW}. In particular 
NP terms coming from S$\tilde\chi$SB effects are expected to give rise to modifications of (gluon and fermion) propagators and fermion-antifermion-gluon vertices, as well as to peculiar gluon-gluon-scalar, fermion-antifermion-scalar, fermion-antifermion-gluon-scalar vertices, etc. Consider, for instance, the small-$b^2$ expansions 
(terms odd in $b$ are excluded by ${\cal L}_{\rm toy}$ symmetries)
\begin{eqnarray}
\hspace*{-.3cm} && \langle O(x,x',...) \rangle\Big{|}^R =
\langle O(x,x',...) \rangle\Big{|}^F -
b^2  \langle O(x,x',...) \!\int\! d^4 z \, [ \, L_6^{\slash\!\!\!\!\!\chi}
+ L_6^{\chi}\,](z) \rangle\Big{|}^F \! 
+ {\rm O}(b^4)  \; , \label{ALCSYM} \\
\hspace*{-.3cm} && O(x,x',...) \!\Leftrightarrow A_\mu^b(x) A_\nu^c(x')[\Phi^\dagger \! \Phi]\!(y) \, ,
\; Q_{L/R}(x) \bar Q_{L/R}(x')[\Phi^\dagger  \! \Phi]\!(y) \, ,
\; Q_{L/R}(x) \bar Q_{L/R}(x') [\Phi^\dagger  \! \Phi](y) A_\mu^b(y') \; , \nonumber
\end{eqnarray}
where the label $|^R$ ($|^F$) means that vev's are taken in the UV-regulated (formal) 
${\cal L}_{\rm toy}$ model and $L_6^{\slash\!\!\!\!\!\chi}$ ($L_6^\chi$) is the $d=6$ 
$\tilde\chi$-breaking (conserving) SLEL operator. Focusing on the O($b^2$) terms from 
$L_6^{\slash\!\!\!\!\!\chi}$ in the r.h.s.\ and looking at the contributions with
just one $\sigma$-propagator, we read off the NP corrections to the gluon-gluon-scalar, $Q_{L/R}$-$\bar Q_{L/R}$-scalar and $Q_{L/R}$-$\bar Q_{L/R}$-gluon-scalar vertices, getting 
\begin{eqnarray}
\hspace{-0.3cm}&&\Delta \Gamma_{AA\Phi}^{bc\,\mu\nu}(k,\ell) \Big{|}^R \!\! =  b^2\Lambda_{s} 
\alpha_s(\Lambda_{s}) \, \frac{\delta^{bc}}{2} \{[k(k+\ell)\delta_{\mu\nu}-k_\mu (k+\ell)_\nu]+[\mu\to\nu]\}  
 \,F_{AA\Phi}\Big{(} \frac{\Lambda_{s}^2}{k^2}, \frac{\Lambda_{s}^2}{\ell^2},
 \frac{\Lambda_{s}^2}{(k+\ell)^2} \Big{)}  \, ,\nonumber\\
\hspace{-0.3cm}&&\Delta \Gamma_{Q \bar Q \Phi}(k,\ell) \Big{|}^R \!\! =  b^2\Lambda_{s} \,
\alpha_s(\Lambda_{s}) \, \frac{i}{2} \gamma_\mu (2k+\ell)_\mu  
 \,F_{Q \bar Q \Phi}\Big{(} \frac{\Lambda_{s}^2}{k^2}, \frac{\Lambda_{s}^2}{\ell^2},
 \frac{\Lambda_{s}^2}{(k+\ell)^2} \Big{)}  \, ,
\label{DELTAGLP} \\
\hspace{-0.3cm}&&\Delta \Gamma_{Q \bar Q A \Phi}^{b,\mu}(k,\ell,\ell') \Big{|}^R \!\! =  b^2\Lambda_{s} \,
\alpha_s(\Lambda_{s}) \, i g_s \lambda^b \gamma_\mu   
 \,F_{Q \bar Q A \Phi}\Big{(} \frac{\Lambda_s^2}{{\rm mom}^2} \Big{)} \, , 
 \quad {\rm mom} \in \{k,\ell,\ell', ... ,\ell'+\ell,k+\ell\} \, .
\nonumber
\end{eqnarray}
$F_{AA\Phi}$, $F_{Q \bar Q \Phi}$ and $F_{Q \bar Q A \Phi}$
are dimensionless functions depending on $\Lambda_{s}^2/{\mbox{mom}}^2$ ratios.
From standard Symanzik arguments, small-$b$ expansions like those in~(\ref{ALCSYM}) 
are expected to be valid for momenta much smaller than the UV-cutoff $b^{-1}$. Like in Wilson LQCD, we assume that the NP effects encoded in~(\ref{ALCSYM})--(\ref{DELTAGLP}) persist up to mom$^2 ={\mbox{O}}(b^{-2})$, and {\it conjecture} 
the asymptotic behaviour 
\[
F_{AA\Phi}\Big{(} \frac{\Lambda_{s}^2}{{\rm mom}^2} \Big{)} 
  \stackrel{{\rm mom}^2\to\infty}\longrightarrow  H_{AA}
\, , \!\quad
F_{Q \bar Q \Phi}\Big{(} \frac{\Lambda_{s}^2}{{\rm mom}^2} \Big{)} 
\stackrel{{\rm mom}^2\to\infty}\longrightarrow H_{Q \bar Q} \, , \!\quad
F_{Q \bar Q A \Phi}\Big{(} \frac{\Lambda_{s}^2}{ {\rm mom}^2 } \Big{)}
\stackrel{{\rm mom}^2\to\infty}\longrightarrow H_{Q \bar Q} \, ,
\]
where $H_{AA}$ and $H_{Q \bar Q}$ are O(1) constants and the last two limits are related by gauge invariance.

Further NP corrections analogous to the ones of eqs.~(\ref{DELTAGLP}) actually
occur for ${\cal L}_{Wil}$-induced vertices. 
Based on the symmetries of the model~(\ref{SULL}), to leading order in $g_s^2$ 
(and $b^2$) these NP terms can be {\it effectively reproduced} in PT
by using {\it ad hoc modified Feynman rules}, namely those that
one infers after adding to the ${\cal L}_{\rm toy}$ Lagrangian the terms 
\begin{equation}
\hspace{-.02cm}
\Delta {\cal L} |_{ad\,hoc} \! = \! 
\frac{b^2}{2} \Lambda_{s} \alpha_s \Big\{\!
{\rm tr}[\Phi^\dagger U \!+ {\rm h.c.} ] 
\Big[ \frac{H_{AA}}{2} {\rm tr}(F F)
+ H_{Q \bar Q} (\bar Q {\slash\!\!\!\!\!\!{\cal D}} Q) \Big]
\! + H_{Wil} [\bar Q_L {\overleftarrow{\cal D}} U {\cal D} Q_R +
{\rm h.c.}]  \!\Big\} \!\!\!
\label{DL-PT-SULL} 
\end{equation}
with $U = \Phi [\Phi^\dagger \Phi]^{-1/2}$. The dimensionless field $U$, 
which transforms like $\Phi$
under $\chi_L \times \chi_R$, must necessarily arise in NP corrections
owing to the exact $\chi_L \times \chi_R$ symmetry of the theory.
\begin{figure}[htbp]
\begin{center}
\includegraphics[scale=0.40,angle=-0]{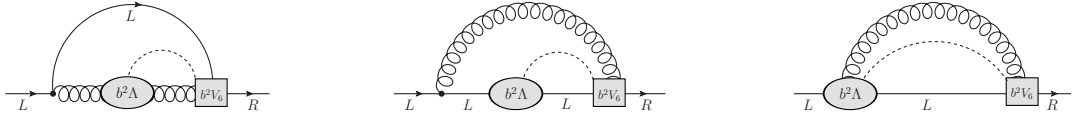}
\end{center}
\caption{Typical lowest order ``diagrams'' giving rise to dynamically generated quark mass
terms (L and R are fermion-helicity labels). The grey blob represents the non-perturbative
$b^2\Lambda_s\alpha_s(\Lambda_s)$ effect in eqs.~(4.6).}
 \label{fig:LEAD-NPMIX-DIAG}
\end{figure}

Formally using PT with modified Feynman rules, one checks that diagrams like those in fig.~\ref{fig:LEAD-NPMIX-DIAG} (the dotted line represents the propagation of a $\sigma \, / \, \pi$ particle) yield a fermion mass O($g^2_s\alpha_s(\Lambda_{s})\Lambda_{s}$). Indeed, for the counting of loop momenta of, say, the central diagram of fig.~\ref{fig:LEAD-NPMIX-DIAG} in the $b\to 0$ limit, one finds a double integral with factors $1/k^2$ and $1/(\ell^2 + m_{\sigma/\pi}^2)$ from the standard gluon and $\sigma \, / \, \pi$ propagators, the factors $\gamma_\mu k_\mu/k^2$ and $\gamma_\nu (k+\ell)_\nu/(k+\ell)^2$ for the quark propagators, a factor $b^2(k+\ell)_\lambda$ from the ${\cal L}_{Wil}$ derivative coupling and a factor $b^2(2k+\ell)_\rho \gamma_\rho \alpha_s(\Lambda_s)\Lambda_s$ from $\Delta \Gamma_{Q \bar Q \Phi}(k,\ell)|^R$. Thus the overall $b^4$ power is compensated by the two-loop integral quartic divergency.

From the NP results we just got 
and the exact $\chi_L\times\chi_R$ invariance  of the theory (see eq.~(\ref{DL-PT-SULL})), we expect the generating functional of 1PI Green functions to display 
a NP mass term of the form~\footnote{An analysis of the WTIs of the $\tilde\chi_L \times \tilde\chi_R$ transformations shows that the term~(\ref{NPMASS}) is RG-invariant~\cite{FRNEW}. This fixes the exact dependence of $C_1$ on $b^{-1}$, and means that $C_1\Lambda_s$ represents the fermion mass value at the UV cutoff.}
\begin{eqnarray}
C_1 \Lambda_s [\bar Q_L U Q_R + \bar Q_R U^\dagger Q_L] \, , \label{NPMASS}
\end{eqnarray}
where to leading order in $g_s^2$ one finds
$C_1 \sim {\rm O}( g_s^2(b^{-1}) \alpha_s(\Lambda_s) )$.

Besides a fermion mass, in the Nambu-Goldstone phase of the model at $\eta=\eta_{cr}$, the term~(\ref{NPMASS}) gives rise to NP $\Phi$-to-fermions and (via fermion loops) $\Phi$-to-gluons couplings, also stemming from the dynamical breaking of the $\tilde\chi_L \times \tilde\chi_R$ invariance.
Nevertheless, the fine tuning $\eta\to\eta_{cr}$, that is crucial to get a fermion mass $\ll v$, is ``natural'' because, besides yielding the restoration of the $\tilde\chi_L \times \tilde\chi_R$ invariance in the Wigner phase, it leads in the Nambu-Goldstone phase to the maximal enhancement of this symmetry that is compatible with its dynamical SSB and related NP effects. 

\section{Conclusions}

We have discussed the possibility that O($g^2_s \alpha_s(\Lambda_{s})\Lambda_{s}$) fermion masses are dynamically generated from an interplay between vanishingly small chirally breaking effects left-over in the ``critical'' theory and power divergencies of loop integrals with the insertion of Wilson-like vertices. We have also shown that it is possible to solve the ``fine tuning'' problem associated with the need of separating  
``large'' from ``small'' mass contributions, in a toy-model where an SU(2) doublet of strongly interacting fermions is coupled to a scalar
via Yukawa and Wilson-like ($\tilde\chi$-breaking) terms.

\acknowledgments  We wish to thank M.\ Testa, G.\ Veneziano and P.\ Weisz for numerous, illuminating  discussions.

\end{document}